\documentclass[a4paper,nobibnotes,prl,nofootinbib,twocolumn,floatfix]{revtex4}

\usepackage{amsmath,amssymb,bbm}
\usepackage{epsfig}
\usepackage{enumerate}

\newcommand{\numero}[1]{\noindent{\bf #1.}~}
\newcommand{\titre}[2]{\numero{#1} {\it #2}\\}

\newcommand{\LambdaQCD}{\Lambda_{\mbox{\footnotesize QCD}}}

\begin{document}
\title{Universality and tree structure of high energy QCD
}

\author{S. Munier}\email{Stephane.Munier@cpht.polytechnique.fr} 
\affiliation{Centre de Physique Th{\'e}orique,
{\'E}cole Polytechnique, 91128 Palaiseau Cedex, France%
\footnote{UMR 7644, unit{\'e} mixte de recherche du CNRS.}}
\author{R. Peschanski}
\email{pesch@spht.saclay.cea.fr}
\affiliation{Service de Physique Th{\'e}orique, CEA/Saclay,
  91191 Gif-sur-Yvette Cedex, France\footnote{%
URA 2306, unit{\'e} de recherche associ{\'e}e au CNRS.}}

\begin{abstract}
Using non-trivial mathematical 
properties of a class of nonlinear evolution equations,
we obtain the universal terms in the asymptotic expansion in rapidity
of the saturation scale and of the unintegrated gluon density 
from the Balitski\u\i-Kovchegov equation.
These terms are independent of the initial conditions and of the details of
the equation. The last subasymptotic terms are new results
and complete the list of all possible universal contributions.
Universality is interpreted in a general qualitative
picture of high energy scattering, in which a scattering process corresponds
to a tree structure probed by a given source.
\end{abstract}

\maketitle


\titre{1}{Introduction}

The energy dependence of hard hadronic cross sections 
is encoded in the evolution 
of the parton distributions with the rapidity $Y$. If the relevant 
scale $Q$ is large enough,
like in the case of deep-inelastic scattering, 
for which $Q$ is the virtuality of 
the photon, the latter can be computed in perturbative QCD.
The Balitski\u{\i}-Fadin-Kuraev-Lipatov (BFKL) equation \cite{BFKL} 
resums the leading
Feynman diagrams, in the form of a linear 
integrodifferential evolution equation for the gluon density. 
Its solution exhibits an exponential growth with $Y$.
At larger $Y$, maybe already attainable at
present colliders \cite{Munier:2001nr}, the BFKL 
equation breaks down because finite parton density
or {\it saturation} effects become sizeable \cite{GLR,MQ}. 
It is replaced by nonlinear evolution equations which aim at describing
such effects \cite{CGC,Balitsky:1995ub,Kovchegov:1999ua}. 
The appropriate equations depend on the physical observables
and on the level of approximation of QCD.
They could also include nonperturbative
contributions \cite{Levin:2003nc}.
Our aim is to show that minimal assumptions on the structure of saturation
equations lead to quite general, i.e. ``universal'', properties of the
solutions.
As a case study, we will consider the
Balitski\u{\i}-Kovchegov (BK) 
equation \cite{Balitsky:1995ub,Kovchegov:1999ua}.


In the large $N_c$ limit, at fixed QCD coupling $\bar\alpha=\alpha_s N_c/\pi$,
and for a homogeneous nuclear target,
the measured gluon density ${\cal N}(k,Y)$ at
transverse momentum
$k$ obeys the BK equation, which reads
\begin{equation}
{\partial_Y}{\cal N}=\bar\alpha
\chi\left(-\partial_L\right){\cal N}
-\bar\alpha\, {\cal N}^2\ .
\label{bk}
\end{equation}
$\partial_L$ stands for the partial derivative with respect to
$\log k^2$ and
the function $\chi$ is the
BFKL kernel
$\chi(\gamma)=2\psi(1)-\psi(\gamma)-\psi(1-\gamma)$.
The nonlinear term in the r.h.s. corresponds to the saturation term.
For its relative simplicity, the BK equation~(\ref{bk})
is the prototype of saturation equations.

However, besides the mathematical challenge posed by
the nonlinear character of saturation equations, there is up to now no 
fully consistent way to include saturation corrections in the
perturbative expansion of QCD. 
Furthermore, 
initial conditions are required
at some low rapidity $Y_0$, 
for all values of $k$:
it is a priori not clear how dependent on them
the solution of Eq.~(\ref{bk}) is, in particular 
for small $k\sim \LambdaQCD$,
which lies beyond the known perturbative domain.
Such sensitivity on the infrared has already been a major
drawback to the BFKL equation \cite{Bartels:1993du}.
One purpose of this Letter is to explain in what respect
the solutions to the BK equation 
are independent of the details of both the nonlinear evolution equation and 
of the initial conditions. 
Let us call this property {\it universality}.\\


\titre{2}{Universal properties of the solutions of the BK equation}

We are going to use general methods for solving nonlinear
evolution equations \cite{brunet,ebert}.
The mathematical results that will be relevant for the BK equation 
require that three main conditions be fulfilled \cite{ebert}.
{\it (i)} ${\cal N}=0$
should be a linearly unstable fixed point, i.e. 
any small fluctuation of $\cal N$ grows to infinity from the
linearized evolution equation;
{\it (ii)} the effect of the nonlinearity is to damp this growth when 
${\cal N}\sim {\cal O}(1)$;
{\it (iii)} the initial condition must be ``sufficiently''
steep at large $k$.
Properties {\it (i)} and {\it (ii)} stem from the parton model in
general and
are verified by the BK equation
in particular. In the case of the BK equation, as noticed in 
Refs.~\cite{Munier:2003vc,Munier:2003sj},
the constraint {\it (iii)} means ${\cal N}(k,Y_0)\ll 1/k^{2\gamma_c}$, 
with $\gamma_c=0.6275\cdots$.
Color transparency of perturbative QCD implies
${\cal N}(k,Y_0)\sim 1/k^2$ at large $k$,
which fulfills this condition.

The class of equations defined by the latter constraints
admit traveling wave solutions\footnote{%
Rigorous mathematical results are available \cite{bramson} 
for the Fisher and Kolmogorov-Petrovsky-Piscounov \cite{KPP} equation
(related to the BK equation in the diffusive approximation
\cite{Munier:2003vc}), and 
have been extended to more general cases \cite{ebert}.}, 
i.e. at large $Y$, ${\cal N}$ is a
uniformly translating front:
\begin{equation}
{\cal N}(k,Y)\sim{\cal N}_\infty\big(\log k^2
-\log Q_s^2(Y)\big)\ .
\label{ninf}
\end{equation}
The mathematical treatment of Eq.~(\ref{bk}) consists in considering
asymptotic expansions in $Y$ of both the saturation scale (related to
the translation
of the traveling wave front with rapidity \cite{Munier:2003vc}) and of the gluon
distribution $\cal N$ (which is the front profile \cite{Munier:2003sj}).
It leads to a hierarchical system of nested ordinary differential
equations.
At each order of the expansion, the integration constants are
determined by a matching procedure \cite{ebert}.
The corresponding general equations can be straightforwardly translated
to the case of the BK equation.

Let us first discuss the saturation scale $Q_s(Y)$. 
In terms of the parameters of the BK equation, 
the systematic expansion\footnote{%
Note that the scale of $Q_s(Y)$ is included in the definition.
Indeed, the saturation scale $Q_s(Y)$ can be defined, e.g., in such a way 
that ${\cal N}(Q_s(Y),Y)={\cal N}_0$, where
${\cal N}_0$ is a given constant.
$Q_s(Y)$ is the position of the traveling wave 
front. All universal terms are independent of ${\cal N}_0$.}  
of $Q_s(Y)$ in $Y$ reads
\begin{multline}
\log Q_s^2(Y)
=\bar\alpha\frac{\chi(\gamma_c)}{\gamma_c} Y-\frac{3}{2\gamma_c}\log Y\\
-\frac{3}{\gamma_c^2}
\sqrt{\frac{2\pi}{\bar\alpha\chi^{\prime\prime}(\gamma_c)}}\frac{1}{\sqrt{Y}}
+{\cal O}(1/Y)\ .
\label{satscal}
\end{multline}
$\gamma_c$ is the solution of the implicit equation
$\chi(\gamma_c)=\gamma_c\chi^\prime(\gamma_c)$, see Ref.~\cite{GLR}.
The successive orders contributing to $\partial \log Q_s^2(Y)/\partial Y$
are displayed in Fig.~1.

A few comments are in order.
The first term in Eq.~(\ref{satscal})
is the dominant evolution term
\cite{Iancu:2002tr}.
The second term \cite{Mueller:2002zm} reflects the delay in the formation 
of the front \cite{Munier:2003sj}.
The third term is a new result for the solution of the BK equation.
These three terms are universal.
They are the only possible
universal terms in this asymptotic expansion.
Indeed, one sees that 
by performing a shift in the rapidity $Y\rightarrow Y+Y_0$ 
in Eq.~(\ref{satscal}), that amounts to modifying the initial conditions, 
one
gets additional $1/Y$ terms, while
the lowest order terms are not affected by the shift.

\begin{figure}[ht]
\begin{center}
\epsfig{file=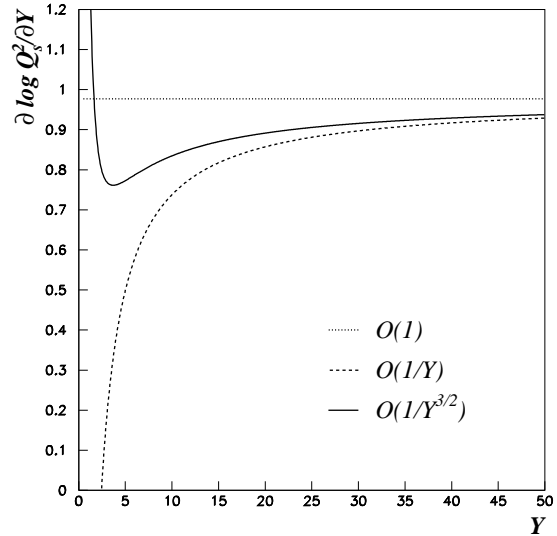,width=8cm}
\caption{{\it The effective dependence $\partial\log{Q_s^2(Y)}/\partial Y$ 
of the saturation scale upon the
  rapidity $Y$.}
$\bar\alpha$ was set to 0.2.
The truncations of the 
asymptotic expansion Eq.~(\ref{satscal}) to order $1$, $1/Y$ and $1/Y^{3/2}$
are shown.
}
\end{center}
\end{figure}

We now turn to the discussion of the properties of the front.
The whole method to get solutions relies on the matching between two 
different expansions of the wave front above the saturation scale:
the so-called ``front interior'' and ``leading
edge'' expansions \cite{ebert}. 
In the ``front interior'', 
the relevant small parameter
is $z=\log(k^2/Q_s^2(Y))/{\sqrt{2\bar\alpha \chi^{\prime\prime}(\gamma_c) Y}}\ll 1$,
i.e. one stands at transverse momenta $k$ near the saturation scale $Q_s(Y)$.
${\cal N}$ is expanded about its asymptotic shape 
${\cal N}_\infty$, see Eq.~(\ref{ninf}). 
In that region, ${\cal N}\sim 1$ and the full nonlinear equation must be 
considered.
In the ``leading edge'' defined by 
$z\sim 1$, the wave front is expanded in powers of $1/\sqrt{Y}$: 
it is the transition region, where
the front forms. 
Going from the front interior to the leading edge,
the
asymptotic front profile ${\cal N}_\infty$ crosses over to \cite{ebert}
\begin{multline}
{\cal N}(k,Y)=C_1\left(\frac{k^2}{Q_s^2(Y)}\right)^{-\gamma_c}
e^{-z^2}
\times\Bigg\{\gamma_c\log [k^2/{Q_s^2(Y)}]\\+C_2
+\left(3-2C_2+\frac{\gamma_c\chi^{(3)}(\gamma_c)}{\chi^{\prime\prime}(\gamma_c)}\right)
z^2\\
-\left({\frac23}\frac{\gamma_c\chi^{(3)}(\gamma_c)}{\chi^{\prime\prime}(\gamma_c)}
+
\frac13{}_2\!F_2\left[1,1;{\scriptstyle\frac52},3;
z^2\right]\right)z^4\\
+6\sqrt{\pi}
\left(1-{}_1\!F_1\left[-{\scriptstyle \frac12},{\scriptstyle \frac32};
z^2
\right]\right)z+{\cal O}(1/\sqrt{Y})
\Bigg\}
\label{front}
\end{multline}
where 
$C_1$ and $C_2$ are two integration constants and ${}_1\!F_1$, ${}_2\!F_2$ are
hypergeometric functions.

The first term can be identified to ${\cal N}_\infty\times
e ^{-z^2}$ near the front, 
and reproduces the known 
results \cite{Iancu:2002tr,Mueller:2002zm,Munier:2003sj}: 
${\cal N}_\infty$ induces so-called 
geometric scaling \cite{Stasto:2000er},
while $e ^{-z^2}$ describes the diffusive formation of the front, 
and induces a specific pattern of geometric scaling violations.
All other terms appearing in the expansion~(\ref{front}) are
new terms that represent the ${\cal O}(1/\sqrt{Y})$
corrections to ${\cal N}_\infty\times
e ^{-z^2}$.
Note that they depend also on the third derivative
$\chi ^{(3)}(\gamma_c)$, 
at variance with the new term of order $1/\sqrt{Y}$
in the expression for the 
saturation scale~(\ref{satscal}).\\

\begin{figure}[ht]
\begin{center}
\epsfig{file=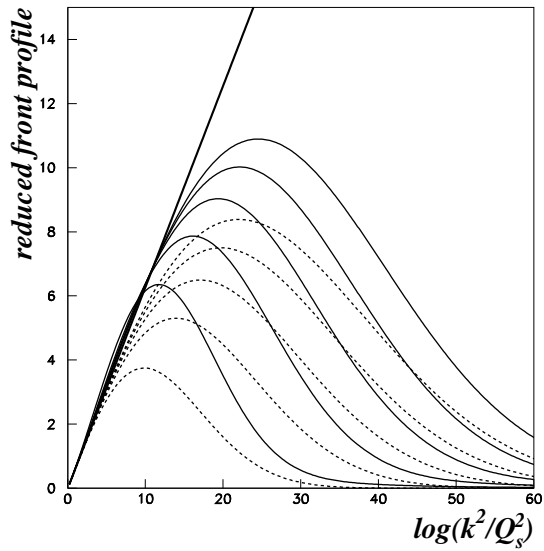,width=8cm}
\caption{{\it 
The reduced front profile ${\cal N}(k,Y)\times (k^2/Q_s^2(Y))^{\gamma_c}$
in the leading edge
}. The curves are for $Y=10$ (lower curves) to $Y=50$ (upper curves).
$\bar\alpha=0.2$ and the constants in Eq.~(\ref{front})
were set to $C_1=1$, $C_2=0$.
The leading edge expansion~(\ref{front}) is truncated at level 
${\cal O}(\sqrt{Y})$ (dashed lines) and 
${\cal O}(1)$ (solid lines).
The reduced asymptotic front, i.e.
${\cal N}_\infty(k,Y)\times (k^2/Q_s^2(Y))^{\gamma_c}$ (see
Eq.~(\ref{ninf})), is  
the straight line.
}
\end{center}
\end{figure}


\titre{3}{Physical interpretation of universality}

Let us discuss the physical picture underlying the universality
properties of a general scattering process in high energy QCD,
 beyond the specific example of the BK equation.
The linear kernels of the saturation equations correspond to 
a developing tree of partons.
The tree structure is generated by successive branching of partons
when the rapidity $Y$ increases, which stems from the nonabelian character of QCD. 
For example, the BFKL kernel is known to correspond to a specific tree of
gluons or, equivalently at large $N_c$, of cascading dipoles \cite{Mueller:1993rr}.

In the course of the parton branching, the partons get more densely packed
together, and thus the color field gets stronger.
The strengh of the field can be qualitatively characterized by the average 
transverse distance $d(Y)$ 
between neighboring partons.

The scattering process corresponds to the interaction of
a source with the partons.
As long as $d(Y)$ is larger than the resolution $1/Q$ of the 
probe, the field seen by the probe is
weak and thus the interaction
merely ``counts'' the partons. This is the dilute regime, in which the
evolution of the observable follows the exponential 
growth of the parton density with $Y$ given by the BFKL equation.
When $d(Y)$ reaches $1/Q$, 
the color field seen by the probe becomes strong,
and the probe can interact with
several partons simultaneously, see Fig.~3. 
This is the saturation transition, and
$d(Y)$ plays the role of the saturation size $1/Q_s(Y)$.\footnote{%
This phenomenon
is conventionally interpreted as
the onset of the black disk due to the filling of the phase space by
partons of size $1/Q$ \cite{GLR}.}
As $d(Y)$ gets even smaller, 
the partons inside the tree may interact and 
recombine. All these finite density or saturation effects
generate nonlinear damping terms in the evolution equations
that slow down the evolution.
Depending on the physical situation,
the evolution of the interaction amplitude in this regime
is e.g. described by the 
BK equation, or by the Balitski\u{\i} \cite{Balitsky:1995ub}
and JIMWLK \cite{CGC} equations which
include more saturation effects.

The structure described here complies with 
the three abovementioned conditions {\it (i), (ii), (iii)} 
for an evolution equation to admit traveling wave solutions.
The previous discussion is general and applies 
to various modifications of the saturation equations.
Therefore, we expect the 
universal results presented here to be valid in general, 
up to the replacement of the relevant parameters related to the
basic tree and which characterize 
the linear evolution kernel.\\

\begin{figure}
\begin{center}
\epsfig{file=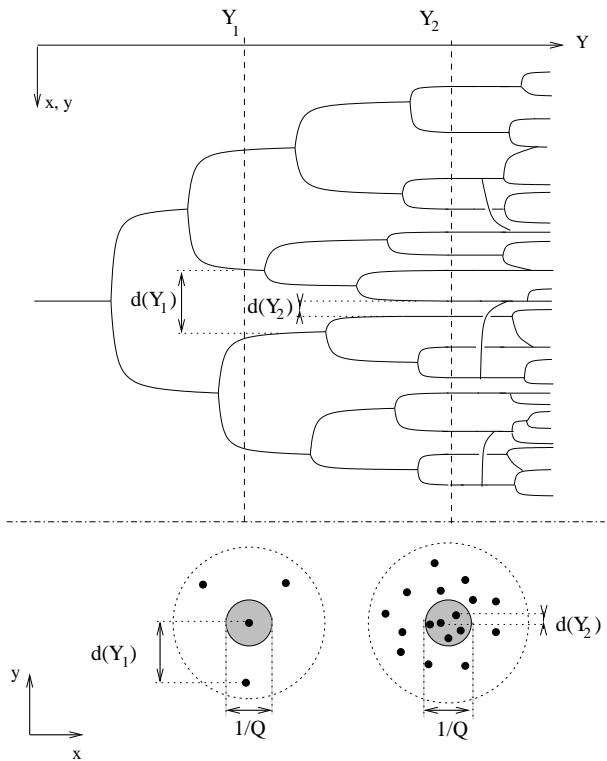,width=8cm}
\end{center}
\caption{{\it The tree structure}.
The parton branching is represented along the rapidity axis (upper
part)
and in the transverse coordinate space at two different rapidities
(lower part). The probe is represented by a shaded disk
of size $1/Q$. At rapidity $Y_1$, the probe counts the partons (linear
regime),
at $Y_2$, the probe counts groups of partons (transition
to saturation). This effect, and also other effects
such as recombination of partons
inside the tree (r.h.s. of the plot), 
generate nonlinear damping terms in the evolution equations.
}
\end{figure}


\titre{4}{Conclusion and discussion}

We have pointed out that the structure of the QCD parton model
is related to general cascading processes possessing specific mathematical
properties.
We have obtained all
universal terms including new ones in
the asymptotic expansion of the saturation scale and 
of the unintegrated gluon
distribution, see Eqs.~(\ref{satscal},\ref{front}).
Universality means independence of the solution
 with respect to the initial conditions after a long enough rapidity
 evolution and dependence of the solution on a few
relevant parameters
obtained from the kernel of the {\it linear} evolution:
$\gamma_c$, $\chi(\gamma_c)$, $\chi^{\prime\prime}(\gamma_c)$ for the
saturation scale, plus
$\chi^{(3)}(\gamma_c)$ for the front.
The results have been obtained from the BK equation, but
are straightforwardly applicable to a whole class of saturation equations.
The saturation scale and the gluon density above the saturation scale
keep exactly the same functional form for the universal terms.

Finally, it is interesting to look for the 
concrete imprints of saturation
on the universal parts of the solution, Eqs.~(\ref{satscal},\ref{front}).
They appear in
two places:
the coefficient of the second term 
of the expansion of the saturation scale
(${\frac32}$ instead of $\frac12$ for BFKL, see Eq.~(\ref{satscal})), and 
the logarithmic factor in ${\cal N}_\infty$ in 
the expression for the asymptotic front (Eq.~(\ref{front})),
that is responsible for the developing linear shape of the reduced
front profile in the leftmost part of the leading edge, see Fig.~2.
This factor appears to be 
a general dynamical property of absorbing walls, near
which a linear gradient develops at long times \cite{ebert}.
The nonlinearities present in the saturation equations
act effectively as an absorbing wall in the results of 
Eqs.~(\ref{satscal},\ref{front}).
This a posteriori justifies 
the treatment of saturation proposed in 
Ref.~\cite{Mueller:2002zm}.

Numerical solutions of the BK \cite{Albacete:2003iq,Rummukainen:2003ns} 
and JIMWLK \cite{Rummukainen:2003ns} equations are available.
The traveling wave behavior of the solution was seen, and the consistency 
with the first two universal
terms in Eq.~(\ref{satscal}) was checked. It would be interesting to include
the newly found term 
in the comparison between the numerical and analytical solutions.
However, the expansion~(\ref{satscal},\ref{front}) 
is an asymptotic series, which is only slowly
converging, see Figs.~1,2.

The question of the phenomenology of saturation is still open.
For realistic phenomenology, one should go beyond leading order BK 
(see e.g. \cite{Triantafyllopoulos:2002nz}) since,
as well-known, the saturation scale predicted by the leading order BK
equation has a too steep rapidity
evolution.

In a forthcoming publication \cite{future}, we take into account some of these
subleading effects by extending the discussion of universality
to the case of the BK equation
with running coupling. The latter falls in a different 
class of nonlinear equations,
but it can be investigated through the same general methods, showing
the power of the mathematical approach.

\begin{acknowledgments}
We thank Edmond Iancu and Dionysis Triantafyllopoulos for their comments.
\end{acknowledgments}


\end{document}